\documentclass[12pt]{article}
\usepackage{bbm}

\newcommand{\eqref}[1]{(\ref{ #1 })}

\usepackage{amssymb}
\usepackage{amsthm}
\usepackage{latexsym}
\def\be{\nopagebreak[3]\begin{equation}}
\def\ee{\end{equation}}
\def\ba{\nopagebreak[3]\begin{eqnarray}}
\def\ea{\end{eqnarray}}

\def\l{\langle}
\def\r{\rangle}

\newcommand{\y}{\hat{y}}              

\newcommand{\py}{{\hat{\pi}^{(y)}}}     
\newcommand{\pyRI}{\hat{\pi}^{(y)R,I}}

\newcommand{\pf}{\hat{\pi}}           
\newcommand{\fluc}[1]{(\Delta #1)^2}      
 
\newcommand{\cre}{\hat{a}^{\dagger}}    

\newcommand{\ann}{\hat{a}}            

\begin{document}
\title{\textbf{The Seeds of Cosmic structure as a door to New Physics}}
\author{Daniel Sudarsky, Instituto de
Ciencias Nucleares,
Universidad Nacional Aut\'onoma de M\'exico,\\
A. Postal 70-543, M\'exico D.F. 04510, M\'exico}
 \vspace{1cm}
\begin{abstract} There is something missing in our understanding of the origin of the seeds of Cosmic Structuture.
 The fact that the  fluctuation spectrum can be extracted  from the inflationary scenario through an analysis that involves quantum field theory in curved space-time,  and that it coincides with the observational data
 has lead to a certain complacency in the community, which prevents the critical analysis of the obscure spots in the derivation. The point is that the inhomogeneity and anisotropy of  our universe  seems to emerge from an exactly homogeneous and isotropic initial state through processes that do not break those symmetries.   This article  gives  a brief recount of the problems faced by the arguments based on established physics, which comprise the point of view held by a large majority of researchers in the field. 
 The conclusion is that  we need  some new physics to be able to fully address the problem.  The article then exposes one avenue that has been used to address the central issue and elaborates on the degree to which, the new approach makes different predictions from the standard  analyses. 
 The approach is inspired on Penrose's proposals that Quantum Gravity might lead to   a real, dynamical collapse of the wave function, a process that we argue has the properties needed to extract us from the theoretical  impasse described above.
  \end{abstract}

\section{Introduction}
\medskip
The last decade has undoubtedly been one of great advances in physical cosmology. 
One of the most important achievements is the precision  measurements of the anisotropies in the CMB\cite{CMB}  together with 
what seems to be their natural explanation within the context of the inflationary scenarios\cite{Guth}.
 There is however a  serious  hole in this seemingly blemish-less suit of the Emperor:  
 The description of our Universe-- or the relevant part thereof- starts\footnote{Here we refer to the era relevant to  the starting point of the analysis that leads to the ``fluctuation spectrum".  In the standard view of inflation, the relevant region of the universe   starts with a Plank regime  containing large  fluctuations of essentially all relevant quantities, but then, a large number of inflationary e-folds leads to an  homogeneous and isotropic universe which is  in fact the starting point of the analysis that takes us to the  
 primordial fluctuation spectrum. One might wish, instead, to regard such fluctuation spectrum as a remnant of the earlier anisotropic and inhomogeneous conditions but then one ends up giving up 
 any pretense that one can explain its  origin and  account for its  specific form.} with an initial condition which is
  homogeneous and isotropic both in the background space-time and in the quantum state that is  supposed to describe the "fluctuations", and it is quite easy to 
 see that the subsequent evolution through dynamics that do not break these symmetries can only lead 
 to an equally homogeneous and anisotropic universe\footnote{ In fact many arguments have been put forward in order to deal with  this issue, that is often phrased in terms of the Quantum to Classical transition -- without focussing on the required 
 concomitant breakdown of homogeneity and isotropy in the state-- the most popular ones associated with the notion of decoherence. These  the alternatives  have been critically discussed in\cite{InflationUS}.}. In fact, if we were to think  in terms of first principles, 
 we would start by acknowledging that the correct description of the  problem at hand would involve a full theory of quantum gravity coupled to a theory of  all the matter quantum  fields, and that there, the issue would be  whether we start with a quantum state that is homogeneous and isotropic or not?. Even if these notions do not make sense within that level of description, a fair question is whether  or not, the inhomogeneities and anisotropies we are interested on, can be  traced to aspects of the description that have no contra-part in the approximation  we are using.  Recall that  such  description involves  the separation of background  vs. fluctuations and thus  must be viewed  only  as an approximation, that allows us to separate the nonlinearities in the system--as well as those  aspects that  are  inherent to  quantum gravity-- from the linear part of problem represented by the fluctuations, which are treated in terms of   linear quantum field theory.  In this sense, we might be tempted to ignore  the problem and view it as something inherent to such approximation.  This would be fine, but we could not argue that we understand the origin of the CMB spectrum, if we view the asymmetries it embodies  as arising from some aspect of the theory we do not rely or touch upon.
 In fact, what we propose in the following treatment  is to bring up one  element or aspect, that  we view as  part of the
 quantum gravity realm, to the forefront in order to modify-- in a minimalistic way-- the semiclassical treatment, that, as we said, we find lacking.
\smallskip
    It is of course not at all  clear that the problem we are  discussing
 should be related to quantum gravity, but
since that is the only sphere of fundamental physics for which  we have so far failed to find 
 a satisfactory  conceptual understanding\footnote{There are of course many open issues in fundamental 
understanding of physics that are not in principle connected with the issue of quantum gravity, however  
it is only in this latter field that the problems seem to be connected with deep conceptual  issues and where 
 one can envision the possibility that their resolution might require a fundamental 
change of paradigm, as  would be the case if we find  we  must modify the laws of quantum mechanics.} we find quite
 natural to associate the two. In this sense we would be  following the ideas of Penrose  regarding the fundamental
 changes,  that he argues\cite{Penrose}, are needed in  quantum mechanics and their connection to quantum gravity.
He argues that quantum gravity might play a role in triggering a real
 dynamical collapse of the 
wave function of systems \cite{Penrose}. His proposals would have a system collapsing whenever the gravitational  interaction 
energy between two alternative 
realizations that appear as superposed in a wave  function of a system reaches a certain threshold which is identified with
 $M_{Planck}$.
The
 ideas  can, in principle, lead to observable effects and, in fact, experiments to test them are currently being contemplated \cite{ExpPenrose}
(although it seems that the available technology can not yet be pushed to the level where actual tests might be expected to become a
 reality soon). We have considered in \cite{InflationUS} a situation for which  there exists already a wealth of empirical information and
 which we have argued
can not be fully understood without  involving some New Physics,  which required  features would  seem to be quite close to Penrose's proposals.

 \section{ The quantum origin of the seeds of cosmic structure}\label{sec_main}
\medskip
 One of the major claimed  successes of Inflationary cosmology is its  reported ability 
to predict the correct spectrum for the primordial density fluctuations that seed the
growth of structure in our Universe. However 
when one thinks about it one immediately notes that there 
is something truly remarkable
 about it, namely that out of an initial situation which is taken to be  perfectly isotropic and homogeneous  
and based on a dynamics that preserves
 those symmetries one ends with a non-homogeneous and non isotropic situation.  Most of our colleagues who have been working in this 
field for a long time would reassure us,  that there is no problem at all by
invoking a variety of arguments. It is noteworthy  that these arguments  would tend to differ in general from one inflationary cosmologist
 to another \cite{Cosmologists}.  Other cosmologists do acknowledge that there seems to be something unclear at this point \cite{Cosmologists2}. In a recent paper \cite{InflationUS},
a critical analysis of such proposals has been carried out indicating that all the existing  justifications fail to be fully
 satisfactory.
 In fact, the  situation  at hand, is quite different from any other situation usually treated using quantum mechanics  as can be seen by noting  that, while in analyzing ordinary situations, quantum  mechanics  offers us, at least one  
self consistent assignment 
at  each time of a state of the Hilbert space to our physical system (we are of course thinking of the Schroedinger picture),  the same is not true for  the standard analysis of the current problem. 
It is well known, that  in certain  instances there might be several 
mutually incompatible assignments of that sort,  as for instance  when contemplating the  two descriptions offered by two different inertial
 observers who
consider a given a specific EPR experiment.
 However,  as  we said, in all other  known cases, one has at least one description available. The reader 
 should try the consideration 
  of  such
 assignment -- of a state at each time -- when presented  with any of the proposed  justifications offered to deal with the issue 
of the
 transition from a
 symmetric universe to a non-symmetric one. The reader will find that  in each case he/she will be asked to accept one of the
 following: i)
our universe was not really 
in that symmetric state (corresponding to the vacuum of the quantum field), ii) our universe is still described by a symmetric state, 
iii) at least at some points in the past the description of the state of our universe could not be done within quantum mechanics, iv) 
quantum mechanics does not correspond to
the full description of  a system at all times,  or v) our own observations of the
 universe mark the transition from a symmetric to an asymmetric state. It should be clear that none of these  represent 
a satisfactory alternative.
In particular, if we want to claim, that we understand the  
evolution of our universe and its structure -- including ourselves -- as the result of the fluctuations of quantum origin in 
the very early stages of 
our cosmology.
Needless is to say that
 none of these options will be explicitly called upon in the arguments one is presented with, however 
 one or more would be
 hidden, perhaps
 in a subtle way, underneath some of the aspects of 
the explanation. For a more thorough discussion we refer the reader to \cite{InflationUS}.

The interesting part of this situation is that one is forced to call upon some novel physical
 process  to fill in  the
 missing or unacceptable part of the justification of the steps that are used to take us from that 
early and symmetric state, to the
 asymmetric state 
of our universe today, or  the state of the universe we photograph when we look at the surface of
 last scattering in the pictures of the CMB.
In \cite{InflationUS} we have considered in this  cosmological context a proposal calling for a 
self induced collapse of the wave function 
along the general 
lines conceived by Penrose, and
have shown that the requirement that one  should obtain results compatible with current observations
 is already sufficient
 to restrict 
in important 
ways some  specific aspects of these novel physics. Thus, when  we consider that the origin of
 such new physics could  be traced to some aspects of quantum gravity, one would be  already in a position of setting
 phenomenological
 constraints on 
at least  this aspect of the quantum 
theory of gravitation.  

In the following we give a short description of this analysis.  The
staring point is as usual the action of a scalar field coupled to
gravity\footnote{We  will be using units where $c=1$ but will keep $\hbar$ (with units of Mass $ M$ times Length  $L$ ), and $G $ ( with units of $ L/M$ ) explicitly throughout the manuscript .The coordinates in the metric $\eta, x^i $ will have units of length $L$  but  the metric components, such as the scale factor $a$  will be dimensionless.  The field  $\phi$ has units of $(M/ L)^{1/2}$,  and the potential $V$ has  units of $M/L^3$}.
\be
\label{eq_action}
S=\int d^4x \sqrt{-g} \lbrack {1\over {16\pi G}} R[g] - 1/2\nabla_a\phi
\nabla_b\phi g^{ab} - V(\phi)\rbrack,
\ee
 where $\phi$ stands for the inflaton or scalar field responsible for inflation and $V$ for the 
inflaton's potential.
 One then splits both, metric and
scalar field into a spatially homogeneous (`background') part and an
inhomogeneous part (`fluctuation'), i.e. $g=g_0+\delta g$,
$\phi=\phi_0+\delta\phi$.

Th equations governing the background unperturbed Friedman-Robertson universe  with line element
$ ds^2=a(\eta)^2\left[- d\eta^2 + \delta_{ij} dx^idx^j\right]$, and the homogeneous scalar field $\phi_0(\eta)$ are, the
scalar field equation, 
\begin{equation}
\ddot\phi_0 + 2 \frac{\dot a}{ a}\dot\phi_0 +
a^2\partial_{\phi}V[\phi] =0, \label{Scalar0}
\end{equation}
and  Friedman's
equation
\begin{equation}
3\frac{\dot a^2}{a^2}=4\pi G  (\dot \phi^2_0+ 2 a^2 V[\phi_0]).
\end{equation}

 The background solution
 corresponds to the standard inflationary cosmology  which written using a conformal time,
 has, during the inflationary era a scale factor
$a(\eta)=-\frac{1}{H_{\rm I} \eta},$
 with $ H_I ^2\approx  (8\pi/3) G V$and with the scalar $\phi_0$ field in the slow roll regime so $\dot\phi_0= - ( a^3/3 \dot a)V'$. This era is supposed to give rise to a reheating period whereby the universe is repopulated with ordinary matter fields, and then to a standard hot big bang cosmology leading up to the present cosmological time. The functional  form  of $a(\eta)$ during these latter periods is  of course different but we will ignore such details on the account that  most of the change in the value of $a$ occurs during the inflationary regime.  The overall normalization of the scale factor will be set so $ a=1$ at the "present  cosmological time". The inflationary regime would end for a value of $\eta=\eta_0$, negative and very small  in absolute terms.
 
 The perturbed metric can be written
\begin{equation}
ds^2=a(\eta)^2\left[-(1+ 2 \Psi) d\eta^2 + (1- 2
\Psi)\delta_{ij} dx^idx^j\right],
\end{equation}
 where $\Psi$  stands for the relevant perturbation and is called
the Newtonian potential.

The perturbation of the scalar field leads to a perturbation of the energy momentum tensor, and
thus Einstein's equations at lowest order lead to
\begin{equation}
\nabla^2 \Psi  = 4\pi G \dot \phi_0 \delta\dot\phi  \equiv s \Gamma 
\label{main3}
\end{equation}
where we introduced the abbreviation $s=4\pi G \dot \phi_0$ and the
quantity $\Gamma$ as the aspect of the field that acts as a source of
the Newtonian Potential,   which for  slow roll approximation  considered here is just 
$\Gamma=\delta\dot\phi$.
Now, write the quantum theory of the field $\delta\phi$.
It is convenient  to consider instead the  field  $y=a \delta \phi$. 
We
consider the field in a box of side $L$, and  decompose the  real
field $y$  into plane waves
\begin{equation}
y(\eta,\vec{x})=\frac{1}{L^{3}} \Sigma_{ \vec k} \left(\ann_k y_k(\eta)
e^{i \vec{k}\cdot\vec{x}}+\cre_{k} \bar y_k(\eta)
e^{-i\vec{k}\cdot\vec{x}}\right),
\end{equation}
where the sum is over the wave vectors $\vec k$ satisfying $k_i L=
2\pi n_i$ for $i=1,2,3$ with $n_i$ integers. 

It is convenient to rewrite the field and momentum operators  as
\begin{equation}
\y(\eta,\vec{x})=
 \frac{1}{L^{3}}\sum_{\vec k}\ e^{i\vec{k}\cdot\vec{x}} \hat y_k
(\eta), \qquad \py(\eta,\vec{x}) =
\frac{1}{L^{3}}\sum_{\vec k}\ e^{i\vec{k}\cdot\vec{x}} \hat \pi_k
(\eta),
\end{equation}
where $\hat y_k (\eta) \equiv y_k(\eta) \ann_k +\bar y_k(\eta)
\cre_{-k}$ and  $\hat \pi_k (\eta) \equiv g_k(\eta) \ann_k + \bar g_{k}(\eta)
\cre_{-k}$
with
\begin{equation}
y^{(\pm)}_k(\eta)=\frac{1}{\sqrt{2k}}\left(1\pm\frac{i}{\eta
k}\right)\exp(\pm i k\eta),\qquad
g^{\pm}_k(\eta)=\pm
i\sqrt{\frac{k}{2}}\exp(\pm i k\eta) . \label{Sol-g} 
\end{equation}

 As we will  be interested in considering a kind of self induced collapse which
 operates in close analogy with  a ``measurement", we proceed to work
 with  Hemitian operators, which in ordinary quantum mechanics are the ones susceptible of direct measurement.
Thus we decompose both $\hat y_k (\eta)$ and $\hat \pi_k
(\eta)$ into their real and imaginary parts $\hat y_k (\eta)=\hat y_k{}^R
(\eta) +i \hat y_k{}^I (\eta)$ and $\hat \pi_k (\eta) =\hat \pi_k{}^R
(\eta) +i \hat \pi_k{}^I (\eta)$ where
\begin{equation}
\hat{y_k}{}^{R,I} (\eta) =
\frac{1}{\sqrt{2}}\left(
 y_k(\eta) \ann_k{}^{R,I}
 +\bar y_k(\eta) \cre{}^{R,I}_k\right) ,\qquad  
\hat \pi_k{}^{R,I} (\eta) =\frac{1}{\sqrt{2}}\left( g_k(\eta)
\ann_k{}^{R,I}
 + \bar g_{k}(\eta) \cre {}^{R,I}_{k} \right).
\end{equation}
We note that the operators $\hat y_k^{R, I} (\eta)$ and $\hat
\pi_k^{R, I} (\eta)$ are therefore hermitian operators.  
Note that the operators corresponding to $k$ and $-k$ are identical in the real
case (and identical up to a sign in the imaginary case).

Next we specify our model of collapse, and follow the field evolution through collapse
to the end of inflation.
  We will assume that the collapse is
somehow analogous to an imprecise measurement of the
operators $\hat y_k^{R, I}
(\eta)$ and $\hat \pi_k^{R, I} (\eta)$ which, as we pointed out are
hermitian operators and thus reasonable observables.  These field
operators contain complete information about
the field (we ignore here for simplicity the relations between the modes $k$ and $-k$).

 Let $|\Xi\rangle$ be any state in the Fock space of
$\hat{y}$. Let us introduce the following quantity:
$d_k^{R,I} = \l \ann_k^{R,I} \r_\Xi.$
Thus  the expectation values of the modes are expressible
as
\begin{equation}
\l {\y_k{}^{R,I}} \r_\Xi = \sqrt{2} \Re (y_k d_k^{R,I}),  \qquad
\l {\py_k{}^{R,I}} \r_\Xi = \sqrt{2} \Re (g_k d_k^{R,I}).
\end{equation}

For the vacuum state $|0\rangle$ we  have of course:
$
\l{\y_k{}^{R,I}}\r_0 = 0, \l\py_k{}^{R,I}\r_0 =0,
$
while their corresponding uncertainties are
\begin{equation}\label{momentito}
\fluc{\y_k {}^{R,I}}_0 =(1/2) |{y_k}|^2(\hbar L^3), \qquad
\fluc{\pf_k {}^{R,I}}_0 =(1/2)|{g_k}|^2(\hbar L^3).
\end{equation}

{\bf The collapse}\newline

Now we will specify the rules according to which collapse happens.
Again, at this point our criteria will be simplicity and naturalness.
Other possibilities do exist, and may lead to different
  predictions. 

What we have to describe is the state $|\Theta\rangle$ after the
collapse. We need to specify 
$d^{R,I}_{k} = \langle\Theta|\ann_k^{R,I}|\Theta\rangle $.
In the vacuum state, $\y_k$ and
$\py_k$ individually are distributed according to Gaussian
distributions centered at 0 with spread $\fluc{\y_k}_0$ and
$\fluc{\py_k}_0$ respectively.  However, since they are mutually
non-commuting, their distributions are certainly not independent.  In
our collapse model, we do not want to distinguish one over the other,
so we will ignore the non-commutativity  and make the following
assumption about the (distribution of) state(s) $|\Theta\rangle$ after
collapse:
\begin{eqnarray}
\l {\y_k^{R,I}(\eta^c_k)} \r_\Theta&=&x^{R,I}_{k,1}
\sqrt{\fluc{\y^{R,I}_k}_0}=x^{R,I}_{k,1}|y_k(\eta^c_k)|\sqrt{\hbar L^3/2},\\
\l {\py_k{}^{R,I}(\eta^c_k)}\r_\Theta&=&x^{R,I}_{k,2}\sqrt{\fluc{\pyRI_k}
_0}=x^{R,I}_{k,2}|g_k(\eta^c_k)|\sqrt{\hbar L^3/2},
\end{eqnarray}
where $x_{k,1},x_{k,2}$ are  selected randomly from within a Gaussian
distribution centered at zero with spread one.
From these equations we  solve for $d^{R,I}_k$.
 Here we must emphasize that our universe, corresponds 
to a single realization of the random variables, and thus  each of the quantities 
$ x^{R,I}{}_{k,1,2}$ has a  single specific value. 
Later, we will see how to make relatively specific predictions, despite  these features.

Next we focus on the expectation value of the quantum
operator which appears in our basic formula
Eq.(\ref{main3}). In the slow roll approximation we have
$\Gamma= a^{-1} \pi^{y}$. Our general approach indicates that, upon
quantization, the above equation turns into
\begin{equation}\nabla^2 \Psi = s \langle\hat\Gamma\rangle. \label{main4}
\end{equation}
Before the collapse occurs, the expectation value on the right hand
side is zero. Let us now determine what happens after the collapse: To
this end, take the Fourier transform of  Eq.(\ref{main4}) and rewrite it
as
\begin{equation}\label{modito}
\Psi_k(\eta)=\frac{s}{k^2}\langle\hat\Gamma_k\rangle_\Theta.
\label{Psi}
\end{equation}

Let us focus now on the slow roll approximation and compute the right
hand side, we note that $\delta\dot\phi=a^{-1}\py$ and hence
 we find
\begin{eqnarray}
\nonumber
\langle\Gamma_k\rangle_\Theta&=&\sqrt{\hbar L^3 k}\frac{1}{2a}F(k), \label{F}
\end{eqnarray}
where
\begin{equation}
F(k) = (1/2) [A_k (x^{R}_{k,1} +ix^{I}_{k,1}) + B_k (x^{R}_{k,2}
+ix^{I}_{k,2})],
\end{equation}
with
\begin{equation}  A_k =  \frac {\sqrt{ 1+z_k^2}} {z_k} \sin(\Delta_k) ; \qquad  B_k
=\cos (\Delta_k) + (1/z_k) \sin(\Delta_k)
\end{equation}
and where  $\Delta_k= k \eta -z_k$ with $ z_k =\eta_k^c
k$.

  Next we turn to the  experimental  results. We will, for the most part,  disregard the changes to
dynamics that happen after re-heating  and due to the transition to
standard (radiation dominated) evolution. The quantity that is measured is ${\Delta T \over T}
(\theta,\varphi)$ which is a function of the coordinates on the
celestial two-sphere which is expressed as $\sum_{lm} \alpha_{lm}
Y_{l,m}(\theta,\varphi)$.  The angular variations of the 
temperature are then identified with the corresponding variations in the
``Newtonian Potential" $ \Psi$, by the understanding that they are the
result of gravitational red-shift in the CMB photon frequency $\nu$ so
${{\delta T}\over T}={{\delta \nu}\over {\nu}} = {{\delta (
    \sqrt{g_{00}})}\over {\sqrt{g_{00}}}} \approx \Psi$.  

 The quantity that is presented
as the result of observations is $OB_l=l(l+1)C_l$ where $C_l =
(2l+1)^{-1}\sum_m |\alpha^{obs}_{lm}|^2 $. The observations indicate
that (ignoring the acoustic oscillations, which is anyway an aspect
that is not being considered in this work) the quantity $OB_l$ is
essentially independent of $l$ and this is interpreted as a reflection
of the ``scale invariance" of the primordial spectrum of fluctuations.

Then, as we noted the  measured quantity is the
``Newtonian potential" on the surface of last scattering: $
\Psi(\eta_D,\vec{x}_D)$,  from where one 
extracts
\begin{equation}
a_{lm}=\int \Psi(\eta_D,\vec{x}_D) Y_{lm}^* d^2\Omega.
\end{equation}
To evaluate the  expected value for the quantity of interest we use (\ref{Psi}) and (\ref{F}) to
write
\begin{equation}
 \Psi(\eta,\vec{x})=\sum_{\vec k}\frac{s  U(k)} {k^2}\sqrt{\frac{\hbar
k}{L^3}}\frac{1}{2a}
 F(\vec{k})e^{i\vec{k}\cdot\vec{x}},
\label{Psi2}
\end{equation}
where we have added the factor $U(k)$ to represent the aspects of
the evolution of the quantity of interest associated with the
physics of period from re-heating to de coupling, which includes among
  others the acoustic oscillations of the plasma.  

 After some algebra we obtain
\begin{eqnarray}
\alpha_{lm}&=&s\sqrt{\frac{\hbar}{L^3}}\frac{1}{2a} \sum_{\vec
k}\frac{U(k)\sqrt{k}}{k^2} F(\vec k)  4 \pi i^l  j_l((|\vec k|
R_D) Y_{lm}(\hat k),\label{alm1}
\end{eqnarray}
where $\hat k$ indicates the direction of the vector $\vec  k$. It is in this
expression that the justification for the use of statistics becomes
clear.  The quantity we are in fact considering is the result of 
the combined contributions of an
ensemble of harmonic oscillators each one contributing with a complex
number to the sum, leading to what is in effect a 2 dimensional random
walk whose total displacement corresponds to the observational
quantity. To proceed further we must evaluate the most likely value
for such total displacement. This we do with the help of the imaginary
ensemble of universes, and the identification of the most likely value
with the ensemble mean vale. Now we
compute the expected magnitude of this quantity. After taking the continuum limit we find, 
\begin{equation}
|\alpha_{lm}|^2_{M. L.} 
=\frac{s^2  \hbar}{2 \pi a^2} \int \frac {U(k)^2
C(k)}{k^4} j^2_l((|\vec k| R_D)  k^3dk, \label{alm4}
\end{equation}
where  
\begin{equation}
C(k)\equiv 1+ (2/ z_k^2) \sin (\Delta_k)^2 + (1/z_k)\sin (2\Delta_k).
\label{ExpCk}
\end{equation}

  The last expression can be made more useful
by changing the variables of integration to $x =kR_D$ leading to
\begin{equation}
|\alpha_{lm}|^2_{M. L.}=\frac{s^2   \hbar}{2 \pi a^2} \int
\frac{U(x/R_D)^2 C(x/R_D)}{x^4}    j^2_l(x) x^3 dx,
\label{alm5}
\end{equation}
which in the exponential expansion regime where $\mu$ vanishes and in
the limit $z_k\to -\infty$ where $C=1$, and taking for simplicity
  $U (k) =U_0$ to be independent of $k$, (neglecting for instance the
  physics that gives rise to the acoustic peaks), we find:
 \begin{equation}
 |\alpha_{lm}|^2_{M. L.}=\frac{s^2  U_0^2  \hbar} {2  a^2}
\frac{1}{l(l+1)} .
\end{equation}

 Now,  since this does not depend on $m$ it
is clear that the expectation of $C_l = (2l+1)^{-1}\sum_m
|\alpha_{lm}|^2 $ is just $|\alpha_{lm}|^2$ and thus the observational
quantity $OB_l=l(l+1)C_l =\frac{s^2 U_0^2 \hbar}{2 a^2} $ independent
of $l$ and in agreement with the scale invariant spectrum obtained in
ordinary treatments and in the observational studies.  

Thus, the predicted value for the  $OB_l$ is,
\begin{equation}\label{resultA}
OB_l= (\pi/3) G\hbar \frac{(V')^2}{V} U_0^2 =
( 2\pi/3)\epsilon \tilde V U_0^2,
\end{equation}
where  we have used the standard definition of the dimensionless
slow roll parameter 
$
\epsilon \equiv (1/2) (M_{Pl}^2/ \hbar) (V'/V)^2
$
which is  normally expected to be rather small and  the dimensionless potential 
$\tilde V \equiv V\hbar^3/M_{Pl}^4$.   Thus,  if $U$ could  be prevented  from becoming
too large during re-heating, the quantity of interest
  would be proportional to $\epsilon$ a possibility
 that was not uncovered  in the standard treatments. That is , the present analysis offers a path to
 get rid of the ``fine tuning problem" for the inflationary
  potential, i.e. even if $ V\hbar^3 \sim M_{Pl}^4$, the temperature
  fluctuations in the CMB could remain  rather small (at the level of $10^{-5}$ as observed  in the CMB).

Now, let us focus on the effect of the finite value of times of
collapse $\eta^c_k$.  That is, we consider the general functional form of
$C(k)$. The first thing we note is that in order to get a reasonable
spectrum there seems to be only one simple option: That $z_k $ be
essentially independent of $k$ that is the time of collapse of the
different modes should depend on the mode's frequency according to
$\eta_k^c=z/k$. This is a rather strong conclusion which could  represent   relevant information about whatever the
mechanism of collapse is. 

 Let us turn next to examine  a simple proposal about the collapse mechanism which following Penrose's ideas is assumed to  be tied to Quantum Gravity, and examine it with the above results in mind.
\section{ A version of  `Penrose's  mechanism' for collapse in the cosmological  setting}
\label{sec_penrose}
\medskip
Penrose has for a long time advocated  that  the collapse of quantum
mechanical wave functions might be a dynamical process independent of observation, and that the
underlying mechanism might be related to gravitational
interaction. More precisely, according to this suggestion, collapse
into one of two quantum
mechanical alternatives would take place when the gravitational
interaction energy between the alternatives exceeds a certain
threshold. In fact, much of the initial motivation for the present
work came from Penrose's ideas and his questions regarding the quantum
history of the universe.

A very naive realization of Penrose's ideas in the present setting
could be obtained as follows: Each mode would collapse by the
action of the gravitational interaction between it's own possible
realizations. In our case, one could estimate the interaction energy
$E_I(k,\eta)$ by considering two representatives of the possible
collapsed states on opposite sides of the Gaussian associated with
the vacuum. Clearly, we must  interpret $\Psi$  as the Newtonian
potential and consequently the right hand side of Eq.
(\ref{main3}), (after a rescaling by $a^{-2}$ to replace  the laplacian expressed in the  
comoving coordinates $x$ to
a laplacian associated with coordinates measuring physical lenght )  should be identified  with  matter density $\rho$.  Therefore, $\rho= a^{-2}\dot\phi_0 \Gamma $, with $\Gamma =\pi^y/a=\dot \delta\phi$. Then we have:
\be\label{GE1}
E_I(\eta)=\int \Psi^{(1)} \rho^{(2)}dV =\int \Psi^{(1)}(x,\eta) \rho^{(2)}(x,\eta)a^3 d^3 x = a
\int \Psi^{(1)}(x,\eta) \dot\phi_0 \Gamma^{(2)}(x,\eta) d^3x. 
\ee 
 Note that in this section we are ignoring the overall sign of this energy which being a  gravitational binding energy would  naturally be negative.
We next express this energy in terms of the Fourier expansion  leading to : 
\be
 E(\eta)= (a/L^6) \Sigma_{k, k'} \Psi_{k}^{(1)}( \eta)
\dot\phi_0 \Gamma^{(2)}_{k'} (\eta) \int  e^{i x (k-k')} d^3x = (a/L^3)\dot\phi_0 \Sigma_{k}
\Psi^{(1)}_{ k}( \eta) \Gamma^{(2)}_{k}(\eta) , 
\ee 
where $(1),(2)$
refer to the two different realizations chosen. Recalling
 that $\Psi_{ k} = ( s/k^2) \Gamma_k$, with $s= 4\pi G\dot\phi_0$,  we find 
  \be
 E(\eta)= 4\pi G (a/L^3)
\dot\phi_0^2\Sigma_{k} (1/k^2)
\Gamma^{(1)}_{k}(\eta) \Gamma^{(2)}_{k}(\eta), 
\ee 
  Using equation (\ref{momentito}), we  estimate $ \Gamma^{(1)}_{k}(\eta) \Gamma_{k}^{(2)}(\eta) $ by 
 $|<\Gamma_k > |^2 = \hbar k L^3 (1/2a)^2$,  and thus we we find 
\be
 E_I(\eta) =  \Sigma_{k}( \pi \hbar  G/ak) (\dot\phi_0)^2.
 \ee
 which can be interpreted as the  sum of  the contributions of each mode to the interaction energy of different alternatives.
 According to all the considerations we have made,  we  view each mode's collapse as occurring independently, so the trigger  for the collapse of mode  $k$ would be, in  accordance to Penrose's ideas, the condition that  this energy  $
E_I(k,\eta)=( \pi \hbar  G/ak) (\dot\phi_0)^2 $ reaches the `one-graviton level',  namely, that  it  equals the value of the Planck Mass $M_p$.  Now we use the specific expressions for the scale factor  $ a=\frac{-1}{\eta H_I}$ and the slow rolling of the background scalar field $\dot \phi_0= (1/3)  (\dot a/a^3 ) V'$ to  find 
\be \label{Emodek}
 E_I(k,\eta)=\frac{\pi \hbar  G}{ 9H_I^2} ( a/k) ( V')^2.
 \ee
Thus the condition determining the time of collapse  $\eta^c_k$ of the mode $k$ is 
\be 
  z_k=\eta^c_k k =\frac{\pi }{9} (\hbar V')^2(H_I M_p)^{-3}=\frac{\epsilon} {8\sqrt {6\pi}}(\tilde V)^{1/2}\equiv  z^c,
 \ee
which is independent of $k$,  and thus,  leads to a roughly  scale invariant spectrum of fluctuations in
accordance with observations. Note that the  energy of mode $k$  in Eq. (\ref{Emodek}) is an increasing function of  conformal time $ \eta$, during the slow roll regime.

We can look closer  into this issue and ask when do the relevant modes collapse?. 
    In order to do this we use the value for  $z^c$
and recall that the time of collapse
 is determined bt $\eta^c_k =z^c/k $, and thus the scale factor at the time of collapse of the modes
with wave number $k$  was 
\be
     a^c_k= (H_I\eta^c_k)^{-1} = (12/\epsilon) k l_p (\tilde V )^{-1}
\ee 
     where $l_p$ stands for  the Planck length.  
 As the  value of the scale factor $a$  at the last scattering surface was $a \approx 10^{-4}$ (recall that the scale factor $a$ has been set so its value today is $1$),  the modes that are relevant to say scales of order $10^{-3}$ of the size of the surface of last scattering (corresponding to a fraction of a degree in today's sky) have $k  \approx 10^{-10} {ly}^{-1}$. 
 
 Thus,  taking $\epsilon \times \tilde V $ to be of order $ 10 ^{-5}$, we have for  those modes     
 $
   a^c_k\approx 10^{-45}
$  
 corresponding to $ N_e =103$ e-folds of total expansion, or something like 80 e-folds before the end of inflation in  standard type of inflationary scenarios. Thus in this scheme inflation must have at least 90 e-folds for it to include  the complete description of  the regime we are considering and to account also for the collapse of  the modes that are of the order of magnitude of  the surface of last scattering itself. The usual requirements of inflation put the lowest bound at something like 60 e-folds of inflation so the present requirement is not substantially stronger.
 
     This result can be directly compared   with the  so called, time of ``Horizon crossing"  $ \eta^H_k $ for mode $k $, corresponding to the physical wavelength reaching the Hubble  distance $ H_I^{-1} $.
        Therefore these latter time is determined from: 
   \be
      a^H_k\equiv   a(\eta^H_k)= k/ (2\pi H_I) = k l_p (3/ 32 \pi^3)^{1/2} (\tilde V)^{-1/2}.
   \ee  
        Thus the ratio of scale factors at collapse and at horizon crossing for a given mode is
$ a^c_k/ a^H_k= (16/\epsilon) (6 \pi^3)^{1/2}  (\tilde V)^{-1/2} $,
     which would ordinarily be a very  large number,  indicating that the collapse time would be much later that the time of  ``Horizon exiting"  or crossing out,  of the corresponding mode.
     
Thus  we find that a naive realization of Penrose's ideas seems, at first sight, to be a good candidate to supply the element that we argued  
is missing in the standard accounts of the emergence of the seeds of cosmic structure from quantum fluctuations during the inflationary regime in the early universe.  However more research along these lines is necessary to find out,  for instance, whether the scheme would imply a second collapse of modes already collapsed, and whether such secondary collapse could disrupt  in a substantial way the 
observational spectrum.

\section{An Alternative Collapse Scheme and the Fine Tuning problem}\label{sec_fineT}
\medskip
  This section should be considered even more speculative that the others because the ideas here proposed have not yet undergone  much substantial checking.  Nevertheless, it seems worthwhile to present it here because it illustrates the power of the new way of looking at some of the relevant issues. We have considered one collapse scheme that seemed very natural. However  there is another scheme that could be considered  even more natural  in light of the point view  explored in the  previous section, that the uncertainties in the matter sources of the gravitational field are the triggers of the collapse. We note that it is only the  conjugate momentum to the field $\pi_k(\eta)$  that acts as source of the ``Newtonian potential" in Eq. (\ref{main3}) and  contributes to the gravitational interaction energy in Eq. (\ref{GE1}), thus it seems natural to assume that it is only this quantity what is subject to the collapse (i.e is only this operator that is subjected to a ``self induced measurement") while the field $y_k(\eta)$  itself is not.
   In this case, the analysis is almost identical:  The collapse is defined by
   \be
\l {\y_k^{R,I}(\eta^c_k)} \r_\Theta=0,\qquad
\l {\py_k{}^{R,I}(\eta^c_k)}\r_\Theta=x^{R,I}_{k}\sqrt{\fluc{\pyRI_k}
_0}=x^{R,I}_{k,2}|g_k(\eta^c_k)|\sqrt{\hbar L^3/2},
\ee
where $x_{k}$ are  selected randomly from within a Gaussian
distribution centered at zero with spread one. Again from these equations we  solve for $d^{R,I}_k$, and proceed as before. The only difference so far is that the function  $C(k)$ containing information about the collapse changes slightly (see \cite{InflationUS})  to: 
\begin{equation}
C'(k)=1+ (1- 1/ z_k^2) \sin (\Delta_k)^2 - (1/z_k)\sin (2\Delta_k).
\label{ExpCk2}
\end{equation}
 Compare the above expression with that  corresponding to the first collapse scheme  Eq. (\ref{ExpCk}).
    However  the point we want to make is that this scheme seems to  be a rather serious candidate to alleviate the fine tuning  problem, that, as we have mentioned in the discussion around Eq. (\ref{resultA}), seems to affect  most inflationary scenarios. The point is that according to quite general analysis, \cite{Garriaga} the quantity\footnote{Note that in contarst with this reference  the analysis here is carried out in terms of the  conformal time, and thus the sligth difference with the experssions in that work.}:
    \be
    \zeta \equiv \Psi + aH \delta \phi/\dot \phi_0
   \ee
   remains constant through the cosmological  evolution even if the equation of state changes, so  it seems natural to expect that in our context the corresponding quantity 
    \be
    \zeta \equiv \Psi + aH \l \delta \phi\r_\Theta /\dot \phi_0= \Psi + H \l y \r_\Theta /\dot \phi_0
   \ee
   would be conserved from immediately after  the collapse (a process through which  the classical equations would not hold) through the reheating and up to the late times associated with the observation,
   (we are essentially relying on Ehrenfest's theorem).  However for the collapse mode we have considered in this section the last term in the equations vanishes just after the collapse  so the value of the Newtonian potential would be that of the estimation we have made  before, and, as indicated in the discussion of the last part of section \ref{sec_main},  this indicates a substantial amelioration of the fine tuning problem.  This   seems a very interesting possibility,  but clearly it must be investigated much more profoundly before any  compelling claims in this regard could be made.

\section{Noteworthy features  of this approach}
\medskip

   The quantities of interest $\alpha_{l,m}$  are now understood as different realizations of the random walk
    described by Eq.(\ref{alm1}), so one can study  their  spreading and  in  clear way compare  the model with the observations.  An interesting research issue would be to estimate how many different modes $k$ effectively  contribute sum, i.e how many steps are  the various  the random walks made of.
    
    Another important observation follows directly from the basic point of view adopted in this analysis:  The source of the  fluctuations that lead to anisotropies and inhomogeneities lies in the quantum uncertainties of the scalar field, which collapse due to some unknown quantum gravitational effect. Once collapsed, these density inhomogeneities and anisotropies feed into the  gravitational degrees of freedom leading to nontrivial  perturbations in the metric functions,  in particular the  so called  newtonian potential.  However, the metric itself is not a source of the quantum  gravitational induced collapse (in following with the equivalence principle the local metric perturbations have  no energy).  Therefore, as the scalar field does not act as a source for the gravitational tensor  modes -- at least not at the lowest order considered here -- the tensor modes can not be excited.  The scheme thus naturally leads to the expectation of a zero-- or at least a  strongly  suppressed-- contribution of the tensor modes to the CMB fluctuation spectrum.
   
  As pointed out at the end  of  section \ref{sec_main} and in section \ref{sec_fineT} this approach also opens new avenues to address the fine tuning problem that  affects most  inflationary models, because one can follow in more detail the  objects that give rise to the anisotropies and inhomogeneities, and by having  the possibility to consider independently the issues relative to formation of the perturbation, and their evolution through the
   reheating era.
         
       And finally and as explicitly exhibited in the previous section, this approach allows us to consider concrete proposals for the physical mechanisms that give rise to the inhomogeneities and anisotropies 
       in the early universe, and confront them with observations.  
       
\section{Conclusions}
\medskip

 We have  reviewed a serious  shortcoming of the inflationary account of the origin of cosmic structure, and have given a brief account of the proposals to deal with them which were 
 first reported in \cite{InflationUS}.
These  lines of inquiry have lead to the  recognition  that something else seems to be needed 
for the whole picture to work and that it  could  be pointing   towards an actual manifestation quantum gravity. 
We have shown that not only 
the issues are susceptible of scientific investigation based on observations, but also that a simple
 account of what is needed, seems to be
 provided by the extrapolation of Penrose's ideas to the cosmological setting.
 
   The scheme exhibits several differences in the predictions as compared with the standard analyses of this problem where the metric and scalar field perturbations are quantized, in particular the suppression of the tensor modes\footnote{See however\cite{Roy} for another scheme which also leads to strong suppression of tensor modes.}.   These predictions can, in principle,  be tested, indicating that an issue that could {\it ab initio} be considered to be  essentially a philosophical problem, leads instead  to truly  physical matters.
   
     In fact, it might well be, that in our  frantic search for physical manifestations of new physics tied to quantum aspects of gravitation,  the most dramatic such  occurrence has been in front of our eyes all this time  and it has  just  been overlooked:  The cosmic structure of the Universe itself.

\section*{Acknowledgments}

\noindent It is a pleasure to acknowledge very  helpful conversations with J. Garriaga, E. Verdaguer and A. Perez. This work was supported
 in part  by DGAPA-UNAM
IN108103 and CONACyT 43914-F grants.

\end{document}